\documentclass[aps,twocolumn,showpacs,pra,twoside,amssymb,amsmath]{revtex4-1}
\usepackage{graphicx}
\usepackage{amssymb}
\usepackage{amsfonts}
\usepackage{bbm}
\usepackage{amsmath}
\usepackage{lmodern}
\usepackage{mathrsfs}
\usepackage[dvipdfm,colorlinks=true, citecolor=blue, urlcolor=blue, linkcolor=blue]{hyperref}
\usepackage{txfonts}

\newcommand{\idol}{\ensuremath{\mathbbm 1}}
\newcommand{\eins}{\ensuremath{\mathbbm 1}}
\newcommand{\tr}{{\rm Tr}}

\begin{document}

\title{Observable estimation of entanglement of formation and quantum discord for\\
 bipartite mixed quantum states}

\author{Chengjie Zhang$^{1}$}
\email{cqtzcj@nus.edu.sg}
\author{Sixia Yu$^{1,2}$}
\author{Qing Chen$^{1,2}$}
\author{C.H. Oh$^{1}$}
\email{phyohch@nus.edu.sg}
\affiliation{$^1$Centre for Quantum Technologies, National University of Singapore, 3 Science Drive 2, 117543, Singapore\\
$^2$Hefei National Laboratory for Physical Sciences at Microscale and Department of Modern Physics, \\ University of Science and Technology of China, Hefei, Anhui 230026, China}

\begin{abstract}
We present observable lower and upper bounds for the entanglement of formation (EOF) and quantum discord (QD), which facilitates estimates of EOF and QD for arbitrary experimental unknown states in finite-dimensional bipartite systems. These bounds can be easily obtained by a few experimental measurements on a twofold copy $\varrho\otimes\varrho$ of the mixed states. Based on our results, we use the experimental measurement data of the real experiment given by Schmid \textit{et al.} [Phys. Rev. Lett. \textbf{101}, 260505 (2008)] to obtain the lower and upper bounds of EOF and QD for the experimental unknown state.
\end{abstract}


\maketitle
\section{Introduction}
In recent years it has become more and more clear that quantum correlations, including quantum entanglement and quantum discord, are basic resources in quantum-information processing \cite{EOFde2}.
Therefore, the quantification of quantum correlations becomes fundamental problems in quantum information theory. However, quantum correlations are not yet fully understood. One of the most important entanglement measures is the entanglement of formation (EOF) \cite{EOFde1}. For pure state $|\varphi\rangle$, it is defined  by $E_F(|\varphi\rangle)=S_V(\varrho_A)$, where $S_V(\varrho)=-\mathrm{Tr}(\varrho\log_2\varrho)$ is the von Neumann entropy and $\varrho_A\equiv\tr_B(|\varphi\rangle\langle\varphi|)$ is the reduced density matrix of subsystem $A$. For mixed state $\varrho$, the EOF is defined by the convex roof,
\begin{equation}\label{}
    E_F(\varrho)=\min_{\{p_i,|\varphi_i\rangle\}}\sum_i p_i E_F(|\varphi_i\rangle),
\end{equation}
where the minimum is taken over all possible ensemble realizations $\varrho=\sum_i p_i |\varphi_i\rangle\langle\varphi_i|$ with $p_i\geq0$ and $\sum_i p_i=1$.

The quantum discord (QD) provides a measure for the quantum correlation beyond entanglement \cite{discord1,discord2}. It is believed that in certain quantum computing tasks there are still quantum advantages in the absence of entanglement. One typical example is the deterministic quantum computation with one qubit \cite{DQC1} in which the QD is proposed to be the reason for the quantum speedup \cite{Datta}. Therefore, the QD has attracted much interest in quantum information theory \cite{Modi,Lang,phase,merge,Markovian1,Markovian2,Markovian3,Markovian4,Markovian5,almost,CP,broadcast,luo,2N,condition1,condition2,condition3,Rahimi,evidence}.
The QD is defined to be \cite{discord1,discord2},
\begin{equation}\label{DA}
    \mathcal{D}_A(\varrho)=\min_{\{E_k\}}\sum_k p_k S_V(\varrho_{B|k})+S_V(\varrho_A)-S_V(\varrho),
\end{equation}
where we perform the positive operator-valued measures (POVMs) $\{E_k\}$ on subsystem $A$ and $p_k=\mathrm{Tr}(E_k\otimes\idol\varrho)$, $\varrho_{B|k}=\tr_A(E_k\otimes\idol\varrho)/p_k$.
The minimum in Eq. (\ref{DA}) can also be taken over all the von Neumann measurements \cite{discord1}, and these two definitions coincide in the case of zero quantum  discord.

Interestingly, there is a close relationship between the EOF and QD, namely the Koashi-Winter relation \cite{KW}:
\begin{eqnarray}\label{KW}
\mathcal{D}_A(\varrho_{AB})-S_V(\varrho_A)+S_V(\varrho_{AB})=E_F(\varrho_{BC}),
\end{eqnarray}
where $E_F(\varrho_{BC})$ is the EOF of $\varrho_{BC}$ and $\varrho_{BC}=\mathrm{Tr}_A(|\psi\rangle\langle\psi|_{ABC})$ with $|\psi\rangle_{ABC}$ being a purification of $\varrho_{AB}$.

Although both the EOF and QD were introduced many years ago, they are notoriously difficult to compute because of the minimization. Until now, only a few analytical results for the EOF and QD have been carried out, such as EOF of two-qubit states \cite{2qubit1,2qubit2}, isotropic states \cite{eof1}, Werner states \cite{werner}, and QD of two-qubit Bell diagonal states \cite{Bell}, rank-2 two-qubit states \cite{rank2} and gaussian states \cite{gaussian}. In order to estimate entanglement and QD, several lower and upper bounds of entanglement measures and QD have been proposed \cite{kai,mintert04,lowerupper,real,lower,upper,eof2,eof3,eof4,estimation}. However, there are only a few bounds for EOF and still no general results of lower and upper bounds of QD for arbitrary finite-dimensional bipartite states. Therefore, general results of analytical lower and upper bounds for the EOF and QD are imminently needed.

In this paper, we present observable lower and upper bounds for EOF and QD, which can allow estimates of EOF and QD for arbitrary experimental unknown states in finite-dimensional bipartite systems. These bounds can be easily obtained by a few experimental measurements on a twofold copy $\varrho\otimes\varrho$ of the mixed states. Based on our results we use the experimental measurement data of the real experiment given in Ref. \cite{real} to obtain the lower and upper bounds of EOF and QD for the experiment unknown state.

\section{Observable lower and upper bounds for entanglement of formation}
Before embarking on our main results, let us briefly review the von Neumann entropy, the linear entropy, and the lower and upper bounds of squared concurrence.
The von Neumann entropy, generalized from the classical Shannon entropy to quantum states, is defined by
\begin{eqnarray}
S_V(\varrho)=-\tr(\varrho\log_2\varrho)=-\sum_i \mu_i\log_2\mu_i\equiv H(\vec{\mu}),
\end{eqnarray}
where $\mu_i$ are eigenvalues of $\varrho$ and $\vec{\mu}$ is the Schmidt vector $\{\mu_1,\mu_2,\cdots,\mu_d\}$. The linear entropy, based on the purity of a quantum state, can be defined as
\begin{eqnarray}
S_L(\varrho)=2(1-\tr\varrho^2)=2(1-\sum_i\mu_i^2),
\end{eqnarray}
where for simplicity we have added a coefficient ``2". The lower and upper bounds of squared concurrence for arbitrary finite-dimensional bipartite states introduced in Refs. \cite{lower,upper},
\begin{eqnarray}
\tr(\varrho\otimes\varrho V_i)\leq[C(\varrho)]^2\leq\tr(\varrho\otimes\varrho K_i)
\end{eqnarray}
where $V_{1}=4(P_{-}^{(1)}-P_{+}^{(1)})\otimes P_{-}^{(2)}$, $V_{2}=4P_{-}^{(1)}\otimes(P_{-}^{(2)}-P_{+}^{(2)})$, $K_{1}=4P_{-}^{(1)}\otimes \eins^{(2)}$ and
$K_{2}=4(\eins^{(1)}\otimes P_{-}^{(2)})$ with $P_{-}^{(i)}$ ($P_{+}^{(i)}$) being the projector on the antisymmetric (symmetric) subspace of the two copies of the $i$th subsystem. It is worth noting that the lower and upper bounds can be expressed as follows,
\begin{subequations}\label{VK}
\begin{eqnarray}
\mathrm{Tr}(\varrho\otimes\varrho V_{1})&=&2(\mathrm{Tr}\varrho^{2}-\mathrm{Tr}\varrho_{A}^{2}),\\
\mathrm{Tr}(\varrho\otimes\varrho V_{2})&=&2(\mathrm{Tr}\varrho^{2}-\mathrm{Tr}\varrho_{B}^{2}),\\
\mathrm{Tr}(\varrho\otimes\varrho K_{1})&=&2(1-\mathrm{Tr}\varrho_{A}^{2}),\\
\mathrm{Tr}(\varrho\otimes\varrho K_{2})&=&2(1-\mathrm{Tr}\varrho_{B}^{2}).
\end{eqnarray}
\end{subequations}
Based on these bounds of the squared concurrence, we shall provide bounds for the EOF.

For simplicity, we use the two denotations $\mathrm{co}(g)$ and $\mathrm{ca}(g)$. Here $\mathrm{co}(g)$ denotes the convex hull of the function $g$, which is the largest convex function that is bounded above by the given function $g$. Conversely, $\mathrm{ca}(g)$ denotes the smallest concave function that is bounded below by the given function $g$. The two denotations have been used to obtain explicit expressions and bounds for the EOF \cite{eof1,eof2,eof3,eof4}.

\textit{Theorem 1.---} For any $m\otimes n$ ($m\geq n$) quantum state $\varrho$, its entanglement of formation $E_F(\varrho)$ satisfies
\begin{eqnarray}\label{EOF}
\mathrm{co}[R_L^{(n)}(\Lambda_5)]\leq E_F(\varrho)\leq\mathrm{ca}[F_U^{(n)}(\Lambda_6)],
\end{eqnarray}
where $\Lambda_5^2=\max\{0,2(\tr\varrho^2-\tr\varrho_A^2),2(\tr\varrho^2-\tr\varrho_B^2)\}$, $\Lambda_6=\min\{2(1-\tr\varrho_A^2),2(1-\tr\varrho_B^2)\}$ (specially $\Lambda_6=\min\{2(1-\tr\varrho_A^2),2(1-\tr\varrho_B^2),1-\tr\varrho_A^2-\tr\varrho_B^2+\tr\varrho^2\}$ for two-qubit states \cite{upper}) and
\begin{eqnarray}
R_L^{(d)}(\lambda)&=&H_2[k\alpha(\lambda)]+k\alpha(\lambda)\log_2k,\label{RL}\\
  \alpha(\lambda)&=&[1+\sqrt{1-(k+1)\lambda^2/(2k)}]/(k+1),\\
F_U^{(d)}(\tau)&=&H_2[\gamma(\tau)]+[1-\gamma(\tau)]\log_2(d-1)\label{FU},\\
\gamma(\tau)&=&[1+\sqrt{(d-1)^2-d(d-1)\tau/2}]/d,
\end{eqnarray}
with $H_2(x)=-x\log_2 x-(1-x)\log_2(1-x)$ being the standard binary entropy function,
$k=\lfloor2/(2-\lambda^2)\rfloor$ where $\lfloor\cdot\rfloor$ denotes the floor function,
$\lambda\in [0,\sqrt{2(d-1)/d}]$ and $\tau\in [0,2(d-1)/d]$.

\textit{Proof.--} We first find the minimal admissible $H(\vec{\mu})$ for a given $\lambda=\sqrt{2(1-\sum_i\mu_i^2)}$, and the maximal admissible $H(\vec{\mu})$ for a given $\tau=2(1-\sum_i\mu_i^2)$. Consider the following two functions,
\begin{eqnarray}
R_L^{(d)}(\lambda)&=&\min_{\vec{\mu}}\Bigg\{H(\vec{\mu})|\lambda=\sqrt{2(1-\sum_{i=1}^{d}\mu_i^2)}\Bigg\},\label{RL2}
\end{eqnarray}
\begin{eqnarray}
F_U^{(d)}(\tau)&=&\max_{\vec{\mu}}\Big\{H(\vec{\mu})|\tau=2(1-\sum_{i=1}^{d}\mu_i^2)\Big\},\label{FU2}
\end{eqnarray}
As shown in Appendix, the minimal $H(\vec{\mu})$ versus $\lambda$ consists of $d-1$ segments and the $k$th segment corresponds to $\vec{\mu}$ in the form $\{t,\cdots,t,1-kt,0,\cdots,0\}$ for $t\in[1/(k+1),1/k]$, and the maximal $H(\vec{\mu})$ versus $\tau$ corresponds to $\vec{\mu}$ in the form $\{t,(1-t)/(d-1),\cdots,(1-t)/(d-1)\}$ for $t\in[1/d,1]$. Therefore, one can obtain  Eqs. (\ref{RL}) and (\ref{FU}).

Suppose that we have already found an optimal decomposition $\sum_j p_j|\psi_j\rangle\langle\psi_j|$ for $\varrho$ to achieve the infimum of $E_F(\varrho)$, then $E_F(\varrho)=\sum_j p_j E_F(|\psi_j\rangle)$ by definition. Since $\mathrm{co}[R_L^{(d)}(\lambda)]$ is a monotonously increasing convex function and satisfies $\mathrm{co}[R_L^{(d)}(\lambda)]\leq H(\vec{\mu})$ for a given $\lambda$, one thus has
\begin{eqnarray}
E_F(\varrho)&=&\sum_j p_j E_F(|\psi_j\rangle)=\sum_j p_j H(\vec{\mu}^j)\geq\sum_j p_j \mathrm{co}[R_L^{(n)}(\lambda^j)]\nonumber\\
&\geq& \mathrm{co}[R_L^{(n)}(\sum_j p_j\lambda^j)]\geq\mathrm{co}[R_L^{(n)}(\Lambda_5)],
\end{eqnarray}
where we have used $\sum_j p_j\lambda^j\geq\Lambda_5$ which has been proved in Ref. \cite{lower}. Meanwhile, since $\mathrm{ca}[F_U^{(d)}(\tau)]$ is a monotonously increasing concave function and satisfies $\mathrm{ca}[F_U^{(d)}(\tau)]\geq H(\vec{\mu})$ for a given $\tau$, one thus has
\begin{eqnarray}
E_F(\varrho)&=&\sum_j p_j E_F(|\psi_j\rangle)=\sum_j p_j H(\vec{\mu}^j)\leq\sum_j p_j \mathrm{ca}[F_U^{(n)}(\tau^j)]\nonumber\\
&\leq& \mathrm{ca}[F_U^{(n)}(\sum_j p_j\tau^j)]\leq\mathrm{ca}[F_U^{(n)}(\Lambda_6)],
\end{eqnarray}
where we have used $\sum_j p_j\tau^j\leq\Lambda_6$ which has been proved in Ref. \cite{upper}.  \hfill  $\blacksquare$

\textit{Remark 1.} Actually, Ref. \cite{eof4} has got a similar result, which is equivalent to $\mathrm{co}[F_L^{(n)}(\Lambda_5^2)]\leq E_F(\varrho)\leq\mathrm{ca}[F_U^{(n)}(\Lambda_6)]$ where $F_L^{(d)}(\tau)$ is defined in the appendix. However, our lower bound $\mathrm{co}[R_L^{(n)}(\Lambda_5)]$ is better than their lower bound $\mathrm{co}[F_L^{(n)}(\Lambda_5^2)]$. Furthermore, we will give analytical results of $\mathrm{co}[R_L^{(n)}(\Lambda_5)]$ and $\mathrm{ca}[F_U^{(n)}(\Lambda_6)]$ in the following, and our analytical results are different from the results shown in \cite{eof4}.

\textit{Remark 2.} The above theorem provides explicit lower and upper bounds of the EOF. Since $\Lambda_5$, $\Lambda_6$ (as well as $\Lambda_1, \cdots, \Lambda_4$ shown in the next section) can be directly measured in experiments using two copies of the state, one can easily obtain the lower and upper bounds of the EOF without quantum state tomography.

As introduced above, $\mathrm{co}(g)$ is the largest convex function that is bounded above by the given function $g$, and $\mathrm{ca}(g)$ denotes the smallest concave function that is bounded below by the given function $g$. From Eq. (\ref{RL}) one can obtain the explicit expression of $\mathrm{co}[R_L^{(d)}(\lambda)]$ which also consists of $d-1$ segments, meanwhile from Eq. (\ref{FU}) the explicit expression of $\mathrm{ca}[F_U^{(d)}(\tau)]$ can be obtained. Therefore, $\mathrm{co}[R_L^{(d)}(\lambda)]$ and $\mathrm{ca}[F_U^{(d)}(\tau)]$ read as follows (the detailed calculations have been given in the appendix),
\begin{widetext}
\begin{eqnarray}
 \mathrm{co}[R_L^{(d)}(\lambda)]&=&\left\{
\begin{array}{ll}
H_2\Big(\frac{1+\sqrt{1-\lambda^2}}{2}\Big), & \lambda \in \big[0,1\big], \\[2mm]
\frac{\log_2(k+1)-\log_2(k)}{\sqrt{\frac{2k}{k+1}}-\sqrt{\frac{2(k-1)}{k}}}\bigg(\lambda-\sqrt{\frac{2(k-1)}{k}}\bigg)+\log_2(k),
& \lambda \in \big[1,\sqrt{\frac{2(d-1)}{d}}\big],
\end{array}%
\right.
\label{lowerbound}\\
 \mathrm{ca}[F_U^{(d)}(\tau)]&=&\left\{
\begin{array}{ll}
H_2[\gamma(\tau)]+[1-\gamma(\tau)]\log_2(d-1), & \tau \in \big[0,\frac{4d-6}{d(d-1)}\big], \\[2mm]
\frac{(d-1)\tau}{2(d-2)}\log_2(d-1)-\frac{(d-1)^2}{d(d-2)}\log_2(d-1)+\log_2d,
& \tau \in \big[\frac{4d-6}{d(d-1)},\frac{2(d-1)}{d}\big],
\end{array}%
\right.
\label{lowerbound2}
\end{eqnarray}
\end{widetext}
where $\gamma(\tau)=[1+\sqrt{(d-1)^2-d(d-1)\tau/2}]/d$ and $k=\lfloor2/(2-\lambda^2)\rfloor$.


\section{Observable lower and upper bounds for quantum discord}
Actually, from the proof of Theorem 1, we can also obtain the lower and upper bounds for QD.


\textit{Theorem 2.---} For any $m\otimes n$ quantum state $\varrho$, the quantum discord satisfies
\begin{equation}\label{lower}
    \mathcal{D}_A(\varrho)\geq R_L^{(m)}(\Lambda_1)-R_U^{(mn)}(\Lambda_2)+\mathrm{co}[R_L^{(n)}(\Lambda_3)],
\end{equation}
where $\Lambda_1^2=2(1-\mathrm{Tr}\varrho_A^2)$, $\Lambda_2^2=2(1-\mathrm{Tr}\varrho^2)$,
$\Lambda_3^2=\max\{0,\ 2(\mathrm{Tr}\varrho_A^2-\mathrm{Tr}\varrho_B^2),\ 2(\mathrm{Tr}\varrho_A^2-\mathrm{Tr}\varrho^2)\}$ and
\begin{eqnarray}
R_U^{(d)}(\lambda)&=&H_2[\beta(\lambda)]+[1-\beta(\lambda)]\log_2(d-1)\label{RU},\\
  \beta(\lambda)&=&[1+\sqrt{(d-1)^2-d(d-1)\lambda^2/2}]/d,
\end{eqnarray}
with $\lambda\in [0,\sqrt{2(d-1)/d}]$.

\textit{Proof.---} We first find the maximal admissible $H(\vec{\mu})$ for a given $\lambda=\sqrt{2(1-\sum_i\mu_i^2)}$. Consider the following function,
\begin{eqnarray}
R_U^{(d)}(\lambda)=\max_{\vec{\mu}}\Bigg\{H(\vec{\mu})|\lambda=\sqrt{2(1-\sum_{i=1}^{d}\mu_i^2)}\Bigg\}.\label{RU2}
\end{eqnarray}
As shown in Appendix, the maximal $H(\vec{\mu})$ versus $\lambda$ corresponds to $\vec{\mu}$ in the form $\{t,(1-t)/(d-1),\cdots,(1-t)/(d-1)\}$ for $t\in[1/d,1]$. Therefore, one can obtain Eq. (\ref{RU}).

In order to find the lower bound of the minimization term in $\mathcal{D}_A(\varrho_{AB})$, we use the Koashi-Winter relation Eq. (\ref{KW}).
Suppose that we have already found an optimal decomposition $\sum_j p_j|\phi_j\rangle\langle\phi_j|$ for $\varrho_{BC}$ to achieve the infimum of $E_F(\varrho_{BC})$, then $E_F(\varrho_{BC})=\sum_j p_j E_F(|\phi_j\rangle)$ by definition. Since $\mathrm{co}[R_L^{(d)}(\lambda)]$ is a monotonously increasing convex function and satisfies $\mathrm{co}[R_L^{(d)}(\lambda)]\leq H(\vec{\mu})$ for a given $\lambda$, one thus has
\begin{eqnarray}
E_F(\varrho_{BC})&=&\sum_j p_j E_F(|\phi_j\rangle)=\sum_j p_j H(\vec{\mu}^j)\geq \sum_j p_j \mathrm{co}[R_L^{(n)}(\lambda^j)]\nonumber\\
&\geq& \mathrm{co}[R_L^{(n)}(\sum_j p_j\lambda^j)]\geq \mathrm{co}[R_L^{(n)}(\Lambda_3)],\label{EF}
\end{eqnarray}
where we have used $\sum_j p_j\lambda^j\geq\Lambda_3$ which has been proved in Ref. \cite{lower}. Moreover, from the definitions in Eqs. (\ref{RL2}) and (\ref{RU2}), we can obtain that
\begin{eqnarray}
S_V(\varrho_A)&\geq& R_L^{(m)}(\Lambda_1),\\
S_V(\varrho_{AB})&\leq& R_U^{(mn)}(\Lambda_2),
\end{eqnarray}
which together with Eq. (\ref{EF}) gives exactly Eq. (\ref{lower}). \hfill  $\blacksquare$

In the following, we shall also present an explicit upper bound for the QD, which is observable as well as the lower bound shown in Theorem 2.


\textit{Theorem 3.---} For any $m\otimes n$ quantum state $\varrho$, the quantum discord satisfies
\begin{equation}\label{upper}
    \mathcal{D}_A(\varrho)\leq R_U^{(m)}(\Lambda_1)-R_L^{(mn)}(\Lambda_2)+\mathrm{ca}[F_U^{(n)}(\Lambda_4)],
\end{equation}
where 
$\Lambda_4=\min\{2(1-\mathrm{Tr}\varrho_B^2),\ 2(1-\mathrm{Tr}\varrho^2)\}$.

\textit{Proof.---} Similar to Theorem 2, we also use the Koashi-Winter relation Eq. (\ref{KW}).
Suppose that we have already found an optimal decomposition $\sum_j p_j|\phi_j\rangle\langle\phi_j|$ for $\varrho_{BC}$ to achieve the infimum of $E_F(\varrho_{BC})$, then $E_F(\varrho_{BC})=\sum_j p_j E_F(|\phi_j\rangle)$ by definition. Since $\mathrm{ca}[F_U^{(d)}(\tau)]$ is a monotonously increasing concave function and satisfies $\mathrm{ca}[F_U^{(d)}(\tau)]\geq H(\vec{\mu})$ for a given $\tau$, one thus has
\begin{eqnarray}
E_F(\varrho_{BC})&=&\sum_j p_j E_F(|\phi_j\rangle)=\sum_j p_j H(\vec{\mu}^j)\leq \sum_j p_j \mathrm{ca}[F_U^{(n)}(\tau^j)]\nonumber\\
&\leq& \mathrm{ca}[F_U^{(n)}(\sum_j p_j\tau^j)]\leq \mathrm{ca}[F_U^{(n)}(\Lambda_4)],\label{EF2}
\end{eqnarray}
where we have used $\sum_j p_j\tau^j\leq\Lambda_4$ which has been proved in Ref. \cite{upper}. Moreover, from the definitions in Eqs. (\ref{RL2}) and (\ref{RU2}), we can obtain that
\begin{eqnarray}
S_V(\varrho_A)&\leq& R_U^{(m)}(\Lambda_1),\\
S_V(\varrho_{AB})&\geq&R_L^{(mn)}(\Lambda_2),
\end{eqnarray}
which together with Eq. (\ref{EF2}) gives exactly Eq. (\ref{upper}). \hfill  $\blacksquare$

\textit{Remark 3.} 
The lower and upper bounds of QD are also valid for the minimum in Eq. (\ref{DA}) being taken over all the von Neumann measurements, since
the QD using POVMs is always smaller than or equal to the one using von Neumann measurements, and $\sum_j p_j\tau^j\leq\Lambda_4$ used in Eq. (\ref{EF2}) still holds for the case using von Neumann measurements.

\section{Examples}
In this section, we will present several examples using our lower and upper bounds of the EOF and QD.

\begin{figure}
\begin{center}
\includegraphics[scale=0.45]{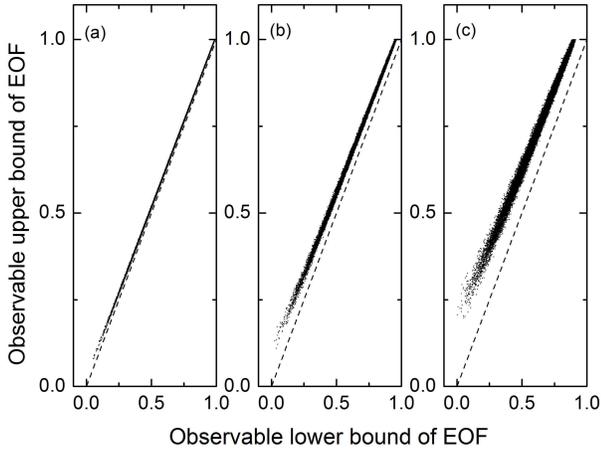}
\caption{Observable upper bound of the EOF versus its observable lower bound for $(2\otimes4)$-dimensional random states with different degrees of mixing: (a) shows weakly mixed states ($0.10\leq\sqrt{1-\tr\varrho^2}\leq0.11$), (b) displays intermediate mixing ($0.20\leq\sqrt{1-\tr\varrho^2}\leq0.21$), and (c) corresponds to strongly mixed states ($0.30\leq\sqrt{1-\tr\varrho^2}\leq0.31$). The dashed lines denote the lower bound.}\label{0}
\end{center}
\end{figure}

\textit{Example 1.---} Considering pure states in $2\otimes n$ systems, one can observe that our lower and upper bounds of EOF and QD coincide. Because for these pure states we have
\begin{eqnarray}
R_L^{(2)}(\Lambda_1)=R_U^{(2)}(\Lambda_1)&=&\mathrm{co}[R_L^{(2)}(\Lambda_5)]=\mathrm{ca}[F_U^{(2)}(\Lambda_6)],\\
R_U^{(2n)}(\Lambda_2)&=&R_L^{(2n)}(\Lambda_2)=0,\\
\mathrm{co}[R_L^{(n)}(\Lambda_3)]&=&\mathrm{ca}[F_U^{(n)}(\Lambda_4)]=0,
\end{eqnarray}
it can be directly obtained that
\begin{equation}\label{}
    E_F(\varrho)=\mathcal{D}_A(\varrho)=R_L^{(2)}(\Lambda_1)=R_U^{(2)}(\Lambda_1).
\end{equation}
Therefore, for the pure sates in $2\otimes n$ systems the EOF and QD can be directly measured using our lower and upper bounds.

\begin{figure}
\begin{center}
\includegraphics[scale=0.45]{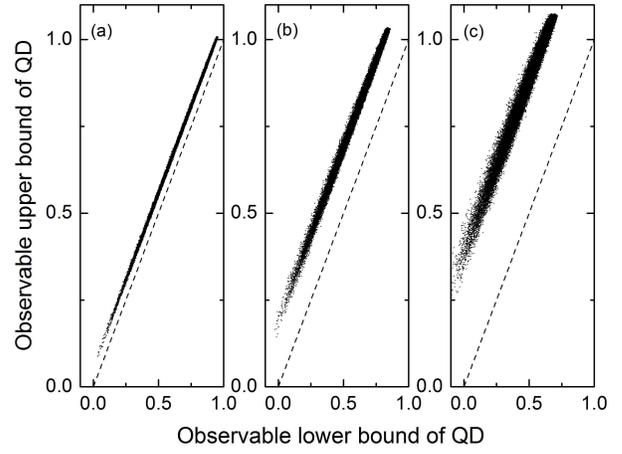}
\caption{Observable upper bound of the QD versus its observable lower bound for the same states used in Fig. \ref{0}.}\label{1}
\end{center}
\end{figure}

\textit{Example 2.---} Let us simulate the lower and upper bounds of the EOF and QD for mixed random states of $(2\otimes4)$-dimensional systems. The mixed random states are generated with different degrees of mixing. Using our theorems, the upper bounds versus the lower bounds of EOF and QD are depicted in Fig. \ref{0} and Fig. \ref{1}, respectively. For weakly mixed states, the lower and upper bounds provide an excellent estimation of EOF and QD; for strongly mixed states, the bounds also provide a region for EOF and QD.

\textit{Example 3.---} Using the experimental measurement data of the real experiment shown in Ref. \cite{real}, one can obtain the lower and upper bounds of the EOF and QD for this experimental unknown two-qubit state $\varrho_{exp}$. Ref. \cite{real} has obtained the following measurement results,
\begin{eqnarray}\label{data}
\wp_{--}=0.208,\ \ \  \wp_{-+}=0.050,\ \ \  \wp_{+-}=0.061,
\end{eqnarray}
where $\wp_{--}=\langle P_{-}^{(1)}\otimes P_{-}^{(2)}\rangle$, $\wp_{-+}$ and $\wp_{+-}$ being defined similarly. Using Eqs. (\ref{VK}) and (\ref{data}), one can calculate that
\begin{eqnarray}
\begin{aligned}
&&\Lambda_1=1.016,  \  \  \  \Lambda_2=0.666, \\
&&\Lambda_3=0.210, \  \  \  \Lambda_4=0.444, \\
&&\Lambda_5=0.795, \  \  \  \Lambda_6=0.832.
\end{aligned}
\end{eqnarray}
Therefore, the lower and upper bounds of the EOF and QD for this experimental unknown state are
\begin{eqnarray}
&&0.715\leq E_F(\varrho_{exp})\leq0.875,\\
&&0.470\leq\mathcal{D}_A(\varrho_{exp})\leq1.023.\label{bounds}
\end{eqnarray}
Although the upper bound in Eq. (\ref{bounds}) gives an invalid bound \cite{invalid}, the lower bound provides an efficient lower bound of QD for the experiment unknown state. The reason for the invalid upper bound comes from the experimental error. It is worth noticing that $\Lambda_1=2(1-\tr\varrho_A^2)$ which should be less than or equal to 1 for two-qubit states. However, in this experiment one has $\Lambda_1=1.016$ because of the experimental error. Therefore, the real upper bound of QD without the experimental error is probably close to and smaller than 1.

\section{Discussion and conclusion}
Let us discuss the realization of the measurements for our lower and upper bounds in real experiments. As shown in Ref. \cite{real}, the antisymmetric projector $P_-$ for two-qubit states takes the particularly simple form $P_-=|\psi^-\rangle\langle\psi^-|$ with $|\psi^-\rangle=(|01\rangle-|10\rangle)/\sqrt{2}$ being the singlet state. It can be measured by several methods using two copies of $\varrho$ in photonic system. For example, Huang \textit{et al.} \cite{Huang} have used a Hong-Ou-Mandel interferometer \cite{Ou} to project two photons onto the singlet state, and Refs. \cite{mintert,real} have used a controlled-NOT gate to distinguish the Bell states, since the controlled-NOT gate can transform the Bell states to separable states and distinguishing separable states is much easier than distinguishing the Bell states.

In this work, we have presented observable lower and upper bounds of the EOF and QD. These bounds can be used to estimate EOF and QD for arbitrary experimental unknown states in finite-dimensional bipartite systems. One can easily obtain these bounds by a few experimental measurements on a twofold copy $\varrho\otimes\varrho$ of the mixed states. Furthermore, based on our results we have used the experimental measurement data of the real experiment given in Ref. \cite{real} to obtain the lower and upper bounds of EOF and QD for the experiment unknown state.

\section*{ACKNOWLEDGMENTS}
This work is supported by the National Research Foundation and Ministry of Education, Singapore (Grant No. WBS: R-710-000-008-271) and NNSF of China (Grant No. 11075227).

\section*{APPENDIX}
In this appendix, we exhibit the details to obtain the expressions of $F_U^{(d)}(\tau)$, $R_L^{(d)}(\lambda)$, $R_U^{(d)}(\lambda)$, $\mathrm{co}[R_L^{(d)}(\lambda)]$ and $\mathrm{ca}[F_U^{(d)}(\tau)]$. The main idea is to illustrate the difference between von Neumann entropy and the linear entropy (or the square root of the linear entropy).

\subsection{Calculation of $F_U^{(d)}(\tau)$, $R_L^{(d)}(\lambda)$, $R_U^{(d)}(\lambda)$}

In the following, we shall seek the highest and lowest von Neumann entropies consistent with a given value of linear entropy. Actually, this problem is equivalent to seek the maximal and minimal $H(\vec{\mu})=-\sum_i \mu_i\log_2\mu_i$ for a given $\tau=2(1-\sum_i\mu_i^2)$, and the later one is a classical problem which has been solved in Refs. \cite{svl1,svl2}. We use $F_U^{(d)}(\tau)$ and $F_L^{(d)}(\tau)$ to denote the maximal and minimal $H(\vec{\mu})$ for a given $\tau$ (although $F_L^{(d)}(\tau)$ has not been used in the main text, we introduce it for the sake of completeness), i.e.,
\begin{eqnarray}
&&F_U^{(d)}(\tau)=\max_{\vec{\mu}}\Big\{H(\vec{\mu})|\tau=2(1-\sum_{i=1}^{d}\mu_i^2)\Big\},\\
&&F_L^{(d)}(\tau)=\min_{\vec{\mu}}\Big\{H(\vec{\mu})|\tau=2(1-\sum_{i=1}^{d}\mu_i^2)\Big\},
\end{eqnarray}
where $d$ denotes the dimension of the Hilbert space.

As shown in Refs. \cite{svl1,svl2}, the maximal $H(\vec{\mu})$ versus $\tau$ corresponds to $\vec{\mu}$ in the form
\begin{eqnarray}\label{form1}
\vec{\mu}=\{t, \frac{1-t}{d-1},\cdots,\frac{1-t}{d-1}\}  \ \ \  \mathrm{for} \  t\in[\frac{1}{d},1]
\end{eqnarray}
with $d-1$ copies of $(1-t)/(d-1)$ and one copy of $t$. Therefore, the maximal $H(\vec{\mu})$ and corresponding $\tau$ are
\begin{eqnarray}
H(t)&=&-t\log_2 t-(1-t)\log_2 \frac{1-t}{d-1},\label{Ht}\\
\tau(t)&=&2[1-t^2-\frac{(1-t)^2}{d-1}].
\end{eqnarray}
In order to show the maximal $H(\vec{\mu})$ versus $\tau$, we need the inverse function of $\tau(t)$. After some algebra, one can obtain that
\begin{eqnarray}\label{t}
t(\tau)=\frac{1}{d}+\sqrt{\frac{(d-1)^2}{d^2}-\frac{(d-1)\tau}{2d}},
\end{eqnarray}
with $\tau\in[0,2(d-1)/d]$. Substituting Eq. (\ref{t}) into Eq. (\ref{Ht}), we can get the expression for $F_U^{(d)}(\tau)$, i.e., Eq. (\ref{FU}) in the main text.

The minimal $H(\vec{\mu})$ versus $\tau$ consists of $d-1$ segments and the $k$th segment corresponds to $\vec{\mu}$ in the form \cite{svl1,svl2}
\begin{eqnarray}\label{form2}
\vec{\mu}=\{t,\cdots,t,1-kt,0,\cdots,0\}  \ \ \  \mathrm{for} \  t\in[\frac{1}{k+1},\frac{1}{k}]
\end{eqnarray}
with $k$ copies of $t$, $d-k-1$ copies of 0 and one copy of $1-kt$. Therefore, the minimal $H(\vec{\mu})$ and corresponding $\tau$ are
\begin{eqnarray}
H(t)&=&-kt\log_2 t-(1-kt)\log_2 (1-kt),\label{Ht2}\\
\tau(t)&=&2[1-kt^2-(1-kt)^2].
\end{eqnarray}
In order to show the minimal $H(\vec{\mu})$ versus $\tau$, we also need the inverse function of $\tau(t)$. After some algebra, one can obtain that
\begin{eqnarray}\label{t2}
t(\tau)=\frac{1}{k+1}+\frac{1}{k+1}\sqrt{1-\frac{(k+1)\tau}{2k}},
\end{eqnarray}
with $\tau\in[0,2(d-1)/d]$. Substituting Eq. (\ref{t2}) into Eq. (\ref{Ht2}), we can get the expression for $F_L^{(d)}(\tau)$, i.e.,
\begin{eqnarray}
F_L^{(d)}(\tau)&=&H_2[k\delta(\tau)]+k\delta(\tau)\log_2k,\label{FL}\\
  \delta(\tau)&=&[1+\sqrt{1-(k+1)\tau/(2k)}]/(k+1),
\end{eqnarray}
with $k=1,\cdots,d-1$. Since $t\in[1/(k+1),1/k]$ for the $k$th segment, we have $\tau\in[2(k-1)/k,2k/(k+1)]$. Therefore, one can obtain $k=\lfloor2/(2-\tau)\rfloor$ where $\lfloor\cdot\rfloor$ denotes the floor function, i.e., $\lfloor x \rfloor$ is the largest integer not greater than $x$.

\begin{figure}
\begin{center}
\includegraphics[scale=0.42]{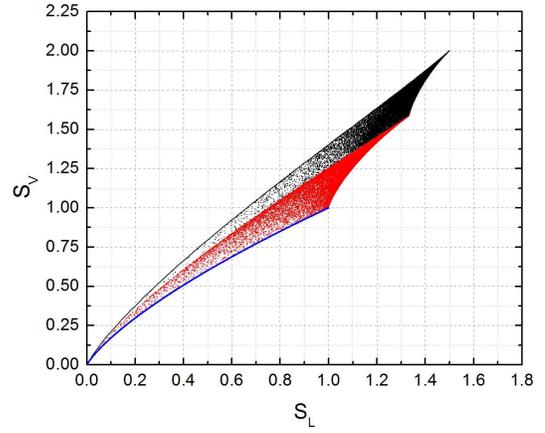}
\caption{(Color online) The linear entropy versus the von Neumann entropy. 50000 dots represent randomly generated states with $d=4$. Rank-2, rank-3, rank-4 states are denoted by the blue, red, black dots, respectively. The upper boundary is a smooth curve, whereas the lower boundary consists of three segments.}\label{2}
\end{center}
\end{figure}

We can simulate the lower and upper boundary of the region in $S_L$ versus $S_V$ plane. 50000 dots for randomly generated states with $d=4$ are displayed in Fig. \ref{2} (a similar figure has also been shown in Ref. \cite{svl2}). The upper boundary corresponds to the function $F_U^{(4)}(\tau)$, and the lower boundary corresponds to $F_L^{(4)}(\tau)$ with three segments.

Similarly, we use  $R_U^{(d)}(\lambda)$ and $R_L^{(d)}(\lambda)$ to denote the maximal and minimal $H(\vec{\mu})=-\sum_i \mu_i\log_2\mu_i$ for a given $\lambda=\sqrt{2(1-\sum_i\mu_i^2)}$, i.e., Eq. (\ref{RU2}) and Eq. (\ref{RL2}) in the main text. Since $\lambda$ is the square root of $\tau$, the forms of $\vec{\mu}$ for the maximal and minimal $H(\vec{\mu})$ versus $\lambda$ are the same as the forms for the maximal and minimal $H(\vec{\mu})$ versus $\tau$, i.e., Eq. (\ref{form1}) and Eq. (\ref{form2}), respectively. Therefore, the maximal $H(\vec{\mu})$ and corresponding $\lambda$ are
\begin{eqnarray}
H(t)&=&-t\log_2 t-(1-t)\log_2 \frac{1-t}{d-1},\label{Htlamda}\\
\lambda(t)&=&\sqrt{2[1-t^2-\frac{(1-t)^2}{d-1}]}.
\end{eqnarray}
and the inverse function of $\lambda(t)$ is
\begin{eqnarray}\label{tlamda}
t(\lambda)=\frac{1}{d}+\sqrt{\frac{(d-1)^2}{d^2}-\frac{(d-1)\lambda^2}{2d}},
\end{eqnarray}
with $\lambda\in[0,\sqrt{2(d-1)/d}]$. Substituting Eq. (\ref{tlamda}) into Eq. (\ref{Htlamda}), we can get the expression for $R_U^{(d)}(\lambda)$, i.e., Eq. (\ref{RU}) in the main text. Furthermore, the minimal $H(\vec{\mu})$ and corresponding $\lambda$ are
\begin{eqnarray}
H(t)&=&-kt\log_2 t-(1-kt)\log_2 (1-kt),\label{Htlamda2}\\
\lambda(t)&=&\sqrt{2[1-kt^2-(1-kt)^2]},
\end{eqnarray}
and the inverse function of $\lambda(t)$ is
\begin{eqnarray}
t(\lambda)=\frac{1}{k+1}+\frac{1}{k+1}\sqrt{1-\frac{(k+1)\lambda^2}{2k}},\label{tlamda2}
\end{eqnarray}
with $\lambda\in[0,\sqrt{2(d-1)/d}]$. Substituting Eq. (\ref{tlamda2}) into Eq. (\ref{Htlamda2}), we can get the expression for $R_L^{(d)}(\tau)$, i.e., Eq. (\ref{RL}) in the main text.

\begin{figure}
\begin{center}
\includegraphics[scale=0.42]{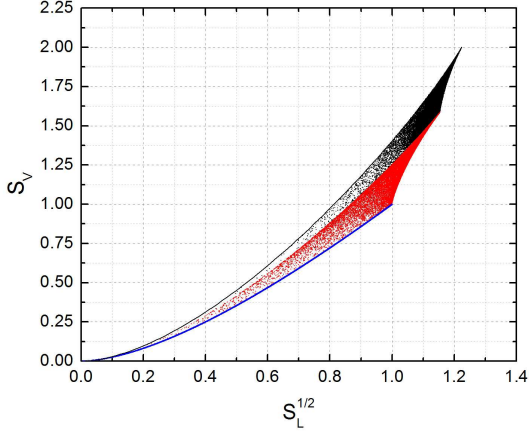}
\caption{(Color online) The square root of linear entropy versus the von Neumann entropy based on the same states used in Fig. \ref{2}.}\label{3}
\end{center}
\end{figure}

Fig. \ref{3} shows the simulation for the lower and upper boundary of the region in $\sqrt{S_L}$ versus $S_V$ plane. The same randomly generated states has been used as Fig. \ref{2}. The upper boundary corresponds to the function $R_U^{(4)}(\lambda)$, and the lower boundary corresponds to $R_L^{(4)}(\lambda)$ with three segments.

\subsection{Calculation of $\mathrm{co}[R_L^{(d)}(\lambda)]$}
We have already obtained the expression for $R_L^{(d)}(\lambda)$ in the above subsection, which comprises $d-1$ segments with $k=1,\cdots,d-1$. In order to obtain $\mathrm{co}[R_L^{(d)}(\lambda)]$, we need to find the largest convex function that is bounded above by $R_L^{(d)}(\lambda)$. It is worth noting that the first segment of $R_L^{(d)}(\lambda)$ (i.e. $k=1$ in Eq. (\ref{RL})) is convex, whereas the rest segments ($k=2,\cdots,d-1$) are concave functions, and one can take Fig. \ref{3} as an example. Therefore, the first segment of $\mathrm{co}[R_L^{(d)}(\lambda)]$ is $R_L^{(d)}(\lambda)$ with $k=1$ itself, i.e., $H_2(1/2+\sqrt{1-\lambda^2}/2)$ for $\lambda\in[0,1]$. The $k$th segment of $\mathrm{co}[R_L^{(d)}(\lambda)]$ ($k=2,\cdots,d-1$) is the line between two points: $(\sqrt{2(k-1)/k},\log_2k)$ and $(\sqrt{2k/(k+1)},\log_2(k+1))$, i.e., $(\lambda-\sqrt{2(k-1)/k})\log_2(1+1/k)/(\sqrt{2k/(k+1)}-\sqrt{2(k-1)/k})+\log_2k$ for $\lambda \in [1,\sqrt{2(d-1)/d}]$. Therefore, the explicit expression of $\mathrm{co}[R_L^{(d)}(\lambda)]$ is Eq. (\ref{lowerbound}).

\subsection{Calculation of $\mathrm{ca}[F_U^{(d)}(\tau)]$}
We have already obtained the expression for $F_U^{(d)}(\tau)$ in the above, which is a smooth function. In order to obtain $\mathrm{ca}[F_U^{(d)}(\tau)]$, we need to find the smallest concave function that is bounded below by $F_U^{(d)}(\tau)$, which is similar to the situation in Ref. \cite{eof3}.

When $d=2$, since $F_U^{(d)}(\tau)$ is a concave function, $\mathrm{ca}[F_U^{(d)}(\tau)]$ is $F_U^{(d)}(\tau)$ itself. When $d\geq3$, we first prove that there is one and only one point $\tau_0$ between 0 and $2(d-1)/d$ such that ${F_U^{(d)}}''(\tau_0)=0$. The second derivative of $F_U^{(d)}(\tau)$ with respect to $\tau$ is
\begin{eqnarray}
{F_U^{(d)}}''(\tau)&=&\bigg[\frac{d}{d\gamma(\tau)-1}\log\frac{(d-1)\gamma(\tau)}{1-\gamma(\tau)}-\frac{1}{\gamma(\tau)(1-\gamma(\tau))}\bigg]\nonumber\\
&&\times(\gamma'(\tau))^2\log_2e,\label{second}
\end{eqnarray}
where $\gamma'(\tau)=-(d-1)/(4d\gamma(\tau)-4)$ and $\gamma''(\tau)=4d(\gamma'(\tau))^3/(d-1)$. From Eq. (\ref{second}) we can obtain ${F_U^{(d)}}''(0)=\lim_{\epsilon\rightarrow 0}{F_U^{(d)}}''(\epsilon)=-\infty$, ${F_U^{(d)}}''(2(d-1)/d)=\lim_{\epsilon\rightarrow 0}{F_U^{(d)}}''(2(d-1)/d-\epsilon)=+\infty$ and $\tau_0$ is the solution of $f(\tau)=g(\tau)$ where
\begin{eqnarray}
f(\tau)&=&\log\frac{(d-1)\gamma(\tau)}{1-\gamma(\tau)},\\
g(\tau)&=&\frac{d\gamma(\tau)-1}{d\gamma(\tau)(1-\gamma(\tau))}.
\end{eqnarray}

\begin{figure}
\begin{center}
\includegraphics[scale=1.2]{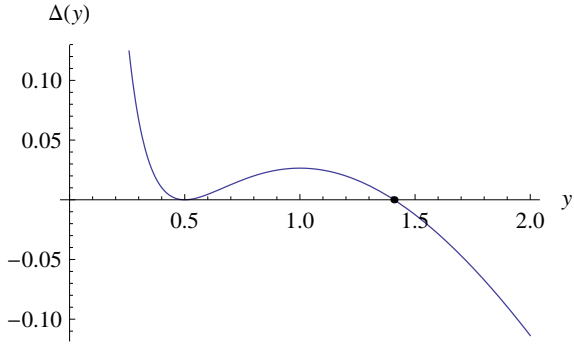}
\caption{(Color online) A typical figure of $\Delta(y)$ versus $y$ when $d=3$. In this case $y_1=1/2$, $y_2=1$, $y_3=2/3$, and $y_0\approx1.40845$.}\label{4}
\end{center}
\end{figure}

For simplicity we define $y=\gamma(\tau)/(1-\gamma(\tau))$ with $y\in(1/(d-1),+\infty)$ because of $\tau\in(0,2(d-1)/d)$. Thus, one can directly consider the functions $f$ and $g$ with respect to $y$,
\begin{eqnarray}
f(y)&=&\log[(d-1)y],\\
g(y)&=&(1-\frac{1}{d})y+(1-\frac{2}{d})-\frac{1}{dy}.
\end{eqnarray}
Therefore, $\tau_0$ corresponds to the solution of $\Delta(y)\equiv f(y)-g(y)=0$. It is worth noting that
\begin{eqnarray}
\Delta(y)&=&\log[(d-1)y]-(1-\frac{1}{d})y-(1-\frac{2}{d})+\frac{1}{dy},\\
\Delta'(y)&=&-\frac{1}{dy^2}+\frac{1}{y}+\frac{1}{d}-1,\\
\Delta''(y)&=&\frac{2-dy}{dy^3}.
\end{eqnarray}
One can directly obtain that $\Delta'(y)=0$ has two roots $y_1=1/(d-1)$ and $y_2=1$, and $\Delta''(y)=0$ has only one root $y_3=2/d$. Moreover, we have $\Delta(y_1)=0$, $\Delta(y_2)>0$ for $d\geq3$, and $\Delta'(y)>0$ when $y\in(1/(d-1),1)$. Therefore, $\Delta(y)$ is a convex function when $y\in(1/(d-1),2/d)$, but $\Delta(y)$ is concave when $y\in[2/d,+\infty)$. Thus, when $d\geq3$ there is only one solution $y_0$ (corresponding to $\tau_0$) for $\Delta(y)=0$ when $y\in(1/(d-1),+\infty)$. A typical figure of $\Delta(y)$ has been shown in Fig. \ref{4} when $d=3$.

The above analysis indicates that for $d\geq3$ $F_U^{(d)}(\tau)$ is concave (i.e. ${F_U^{(d)}}''(\tau)<0$) near $\tau=0$ but convex (i.e. ${F_U^{(d)}}''(\tau)>0$) near $\tau=2(d-1)/d$. In order to obtain $\mathrm{ca}[F_U^{(d)}(\tau)]$, we should solve the following equations: Let
\begin{equation}\label{}
E_{\mathrm{line}}(\tau)=a\bigg[\tau-\frac{2(d-1)}{d}\bigg]+\log_2 d
\end{equation}
be the line crossing through the point $(2(d-1)/d,\log_2 d)$. We solve
\begin{eqnarray}
E_{\mathrm{line}}(\tau)&=&F_U^{(d)}(\tau),\\
\frac{dE_{\mathrm{line}}(\tau)}{d\tau}&=&\frac{dF_U^{(d)}(\tau)}{d\tau}=a,
\end{eqnarray}
for $a$ and $\tau$. The solution to the equations is unique: $\tau=(4d-6)/(d(d-1))$ and $a=(d-1)/(2d-4)\log_2(d-1)$. Therefore, for $d\geq3$ $\mathrm{ca}[F_U^{(d)}(\tau)]$ is $F_U^{(d)}(\tau)$ itself when $\tau\in[0,(4d-6)/(d(d-1))]$ and the line $E_{\mathrm{line}}$ when $\tau\in[(4d-6)/(d(d-1)),2(d-1)/d]$, respectively.


\begin{thebibliography}{99}

\bibitem{EOFde2} M. Horodecki, Quantum Inf. Comput. \textbf{1}, 3 (2001); D. Bru{\ss}, J. Math. Phys. (N.Y.) \textbf{43}, 4237 (2002); M. B. Plenio and S. Virmani, Quantum Inf. Comput. \textbf{7}, 1 (2007); R. Horodecki, P. Horodecki, M. Horodecki and K. Horodecki, Rev. Mod. Phys. \textbf{81}, 865, (2009).

\bibitem{EOFde1} C. H. Bennett, D. P. DiVincenzo, J. A. Smolin, and W. K. Wootters, Phys. Rev. A \textbf{54}, 3824 (1996).

\bibitem{discord1} H. Ollivier and W. H. Zurek, Phys. Rev. Lett. \textbf{88}, 017901 (2001).

\bibitem{discord2} L. Henderson and V. Vedral, J. Phys. A \textbf{34}, 6899 (2001).

\bibitem{DQC1} E. Knill and R. Laflamme, Phys. Rev. Lett. \textbf{81}, 5672 (1998).

\bibitem{Datta} A. Datta, A. Shaji, and C. M. Caves, Phys. Rev. Lett. \textbf{100}, 050502 (2008).

\bibitem{Modi} K. Modi, T. Paterek, W. Son, V. Vedral, and M. Williamson, Phys. Rev. Lett. \textbf{104}, 080501 (2010).

\bibitem{Lang} M. D. Lang and C. M. Caves, Phys. Rev. Lett. \textbf{105}, 150501 (2010).

\bibitem{phase} T. Werlang, C. Trippe, G. A. P. Ribeiro, and G. Rigolin, Phys. Rev. Lett. \textbf{105}, 095702 (2010).

\bibitem{merge} D. Cavalcanti, L. Aolita, S. Boixo, K. Modi, M. Piani, A.Winter, Phys. Rev. A \textbf{83}, 032324 (2011); V. Madhok and A. Datta,  \textit{ibid.} \textbf{83}, 032323 (2011).

\bibitem{Markovian1} T. Werlang, S. Souza, F. F. Fanchini, and C. J. Villas Boas, Phys. Rev. A \textbf{80}, 024103 (2009).

\bibitem{Markovian2} J. Maziero, L. C. C\'{e}leri, R. M. Serra, and V. Vedral, Phys. Rev. A \textbf{80}, 044102 (2009).

\bibitem{Markovian3} F. F. Fanchini, T. Werlang, C. A. Brasil, L. G. E. Arruda, and A. O. Caldeira, Phys. Rev. A \textbf{81}, 052107 (2010); F. F. Fanchini, M. F. Cornelio, M. C. de Oliveira, and A. O. Caldeira,  \textit{ibid.} \textbf{84}, 012313 (2011).

\bibitem{Markovian4} J. Maziero, T. Werlang, F. F. Fanchini, L. C. Celeri, and R. M. Serra, Phys. Rev. A \textbf{81}, 022116 (2010).

\bibitem{Markovian5} J. S. Xu, X. Y. Xu, C. F. Li, C. J. Zhang, X. B. Zou, and G. C. Guo, Nature Commun. \textbf{1}, 7 (2010); J. S. Xu, C. F. Li, C. J. Zhang, X. Y. Xu, Y. S. Zhang, and G. C. Guo, Phys. Rev. A \textbf{82}, 042328 (2010).

\bibitem{almost} A. Ferraro, L. Aolita, D. Cavalcanti, F. M. Cucchietti, and A. Ac\'{i}n, Phys. Rev. A \textbf{81}, 052318 (2010).

\bibitem{CP} A. Shabani and D. A. Lidar, Phys. Rev. Lett. \textbf{102}, 100402 (2009); A. Shabani and D. A. Lidar, Phys. Rev. A \textbf{80}, 012309 (2009);
C. A. Rodr\'{\i}guez-Rosario \textit{et al.}, J. Phys. A \textbf{41}, 205301 (2008).

\bibitem{broadcast} M. Piani, P. Horodecki, and R. Horodecki, Phys. Rev. Lett. \textbf{100}, 090502 (2008).

\bibitem{luo} S. Luo and W. Sun, Phys. Rev. A \textbf{82}, 012338 (2010).

\bibitem{2N} B. Bylicka, and D. Chru\'{s}ci\'{n}ski, Phys. Rev. A \textbf{81}, 062102 (2010).

\bibitem{condition1} B. Daki\'{c}, V. Vedral, and \v{C}. Brukner, Phys. Rev. Lett. \textbf{105}, 190502 (2010).

\bibitem{condition2} A. Datta, arXiv:1003.5256.

\bibitem{condition3} L. Chen, E. Chitambar, K. Modi, and G. Vacanti, Phys. Rev. A \textbf{83}, 020101(R) (2011).

\bibitem{Rahimi} R. Rahimi and A. SaiToh, Phys. Rev. A \textbf{82}, 022314 (2010).

\bibitem{evidence} A. Datta and G. Vidal, Phys. Rev. A \textbf{75}, 042310 (2007).

\bibitem{KW} M. Koashi and A. Winter, Phys. Rev. A \textbf{69}, 022309 (2004).

\bibitem{2qubit1} S. Hill and W. K. Wootters, Phys. Rev. Lett. \textbf{78}, 5022 (1997).

\bibitem{2qubit2} W. K. Wootters, Phys. Rev. Lett. \textbf{80}, 2245 (1998).

\bibitem{eof1} B. M. Terhal and K. G. H. Vollbrecht, Phys. Rev. Lett. \textbf{85}, 2625 (2000).

\bibitem{werner} K. G. H. Vollbrecht and R. F. Werner, Phys. Rev. A \textbf{64}, 062307 (2001).

\bibitem{Bell} S. Luo, Phys. Rev. A \textbf{77}, 042303 (2008).

\bibitem{rank2} L. X. Cen, X. Q. Li, J. Shao, and Y.J. Yan, Phys. Rev. A \textbf{83}, 054101 (2011).

\bibitem{gaussian} P. Giorda and M. G. A. Paris, Phys. Rev. Lett. \textbf{105}, 020503 (2010); G. Adesso and A. Datta, \textit{ibid.} \textbf{105}, 030501 (2010).

\bibitem{kai} K. Chen, S. Albeverio, and S.-M. Fei, Phys. Rev. Lett. \textbf{95}, 040504 (2005).

\bibitem{mintert04} F. Mintert, M. Ku\'{s}, and A. Buchleitner, Phys. Rev. Lett. \textbf{92}, 167902 (2004).

\bibitem{lowerupper} S. Yu, C. Zhang, Q. Chen, and C.H. Oh, arXiv:1102.1301.

\bibitem{real} C. Schmid, N. Kiesel, W. Wieczorek, H. Weinfurter, F. Mintert, and A. Buchleitner, Phys. Rev. Lett. \textbf{101}, 260505 (2008).

\bibitem{lower} F. Mintert and A. Buchleitner, Phys. Rev. Lett. \textbf{98}, 140505 (2007); L. Aolita, A. Buchleitner, and F. Mintert, Phys. Rev. A \textbf{78}, 022308 (2008).

\bibitem{upper} C. J. Zhang, Y. X. Gong, Y. S. Zhang, and G. C. Guo, Phys. Rev. A \textbf{78}, 042308 (2008).

\bibitem{eof2} K. Chen, S. Albeverio, and S.-M. Fei, Phys. Rev. Lett. \textbf{95}, 210501 (2005).

\bibitem{eof3} S. M. Fei and X. Li-Jost, Phys. Rev. A \textbf{73}, 024302 (2006).

\bibitem{eof4} M. Li and S.-M. Fei, Phys. Rev. A \textbf{82}, 044303 (2010).

\bibitem{estimation} G. Brida \textit{et al.}, Phys. Rev. Lett. \textbf{104}, 100501 (2010); Phys. Rev. A \textbf{83}, 052301 (2011).

\bibitem{invalid} For two-qubit states, the quantum discord should be less than or equal to 1. The upper bound 1.02 gives no information of the quantum discord.

\bibitem{Huang} Y.-F. Huang \textit{et al.}, Phys. Rev. A \textbf{79}, 052338 (2009).

\bibitem{Ou} C.-K. Hong, Z.-Y. Ou, and L. Mandel, Phys. Rev. Lett. \textbf{59}, 2044 (1987).

\bibitem{mintert} S. P. Walborn \textit{et al.}, Nature (London) \textbf{440}, 1022 (2006); S. P. Walborn, P. H. Souto Ribeiro, L. Davidovich, F. Mintert, and A. Buchleitner, Phys. Rev. A \textbf{75}, 032338 (2007).

\bibitem{svl1} P. Harremo\"{e}s and F. Tops{\o}e, IEEE Trans. Inform. Theory, \textbf{47}, 2944 (2001).

\bibitem{svl2} T.-C. Wei, K. Nemoto, P. M. Goldbart, P. G. Kwiat, W. J. Munro, and F. Verstraete, Phys. Rev. A \textbf{67}, 022110 (2003).

\end{thebibliography}
\end{document}